\documentclass[12pt]{article}

\usepackage{amsmath, amsfonts, amssymb}
\usepackage{graphicx,psfrag,epsf}
\usepackage{enumerate}
\usepackage{natbib}
\usepackage{xcolor}
\usepackage[normalem]{ulem}
\usepackage{changepage}
\newtheorem{prop}{Proposition}

\newtheorem{myproof}{Proof}
\newtheorem{myassump}{Assumption}
\usepackage{caption}
\usepackage{bm}
\usepackage{natbib}

\usepackage[plain,noend]{algorithm2e}

\begin{document}

\title{\bf Exact Confidence Intervals for Linear Combinations of Multinomial Probabilities}
  
\author{Katherine A. Batterton\footnote{The views expressed in this article are those of the author and do not reflect the official policy or position of the United States Air Force, Department of Defense, or the U.S. Government.} \\
Department of Mathematics and Statistics \\ Air Force Institute of Technology \\ Wright Patterson AFB, OH 45433, USA \\ (katherine.batterton@gmail.com) \\
\\

Christine M. Schubert$^{*}$ \\ Department of Mathematics and Statistics \\ Air Force Institute of Technology \\ Wright Patterson AFB, OH 45433, USA \\ (Christine.Schubert@afit.edu) \\

\\
Richard L. Warr  \\  Department of Statistics \\ Brigham Young University \\ Provo, UT 84602, USA \\ (warr@stat.byu.edu) }

\maketitle
\thispagestyle{empty}

\begin{abstract}
Linear combinations of multinomial probabilities, such as those resulting from contingency tables, are of use when evaluating classification system performance. While large sample inference methods for these combinations exist, small sample methods exist only for regions on the multinomial parameter space instead of the linear combinations. However, in medical classification problems it is common to have small samples necessitating a small sample confidence interval on linear combinations of multinomial probabilities. Therefore, in this paper we derive an exact confidence interval, through the use of fiducial inference, for linear combinations of multinomial probabilities. Simulation demonstrates the presented interval's adherence to exact coverage. Additionally, an adjustment to the exact interval is provided, giving shorter lengths while still achieving better coverage than large sample methods. Computational efficiencies in estimation of the exact interval are achieved through the application of a fast Fourier transform and combining a numerical solver and stochastic optimizer to find solutions. The exact confidence interval presented in this paper allows for comparisons between diagnostic methods  previously unavailable, demonstrated through an example of diagnosing chronic allograph nephropathy in post kidney transplant patients.   
\end{abstract}

\noindent%


\section{Introduction}

\label{sec:intro}
This paper presents an exact confidence interval for the linear combination of multinomial probabilities through the use of fiducial inference. By exact, we mean a confidence interval with a confidence coefficient greater than or equal to the stated confidence level \citep{AgrCoull}. While the interval presented is appropriate for any sample size, it is especially useful for small sample scenarios for which existing methods are not well suited. This is because confidence as applied to linear combinations of multinomial or binomial probabilities has been directed towards the asymptotic properties of a large sample. ~\cite{Gold1963} developed a confidence interval for a generic linear combination of multinomial probabilities using large sample theory, which was then applied to developing the confidence interval of linear functions of the transition probabilities in finite Markov chains. ~\cite{Goodman1964} expanded this concept specifically for contrasts of multinomial probabilities, also under large sample theory. 

\textcolor{black}{When }\textcolor{black}{the desired inference is simultaneous confidence intervals on the multinomial parameters instead of linear combinations of the parameters, large and small sample methods exist. For large samples see~\citet[][]{Gold1963, QH64, LGoodman65, FS87, SG95}, and for small samples \citet{chaf2009} developed a confidence region around multinomial parameters. Simultaneous inference on the multinomial parameter space does not efficiently or directly translate to the linear combinations of those parameters, and therefore the work of this paper fills a gap by providing an exact confidence interval on linear combinations of multinomial probabilities.}

\textcolor{black}{Under certain conditions,} the fiducial argument from~\citet{Fisher1} \textcolor{black}{has proven} useful for deriving \textcolor{black}{approximate and} exact small-sample inference methods. \textcolor{black}{Recent implementation by several authors demonstrate its use}~\citep[see, for example,][]{Xinmin, Krish, Zhao2012, Hannig2009, Batterton2}. Possibly the best known example of the implementation of the fiducial argument is presented by~\citet{CP34} for constructing an exact confidence interval around a binomial proportion.  To develop confidence intervals around linear combinations of multinomial probabilities, the fiducial method also proves useful. \textcolor{black}{While the fiducial approach does not extend universally for multi-parameter distributions \citep{Pedersen1978, Zabell},} in this work, the statistic of interest is a linear function of discrete random variables representing a projection into the one-dimensional real space. \textcolor{black}{Therefore, with careful construction, a confidence interval on the linear combinations of multinomial probabilities with exact frequentist coverage is developed using the fiducial approach.}

While there are many uses for inference on the linear combination of multinomial probabilities, the motivation for this paper comes from quantifying the performance of a medical diagnostic test. There have been many proposed metrics to quantify the performance of a diagnostic test~\citep[]{Zou2013, Unal2017}. The most common metric to estimate correct classification of a diagnostic test is the Youden index, although additional utility has been recently explored with the Bayes cost metric which minimizes the misclassification of the diagnostic test~\citep{RefWorks7, Skaltsa2010, RefWorks6, RefWorks22, Batterton1}.  The advantage of the Bayes cost metric resides in its greater flexibility in weighting probabilities associated with misclassifications when there are more than two diagnostic outcomes~\citep{RefWorks22, Batterton1}. However, by either formation, the criterion of examining correct or mis-classifications is expressed as linear combinations of classification probabilities; probabilities fundamentally distributed as binomial or multinomial, depending on the number of diagnostic outcome classes.  When new diagnostic tests are explored or evaluated against standard tests, statistical comparisons based on classification performance are of great use. For such statistical inference, large sample theory may be applied. However, it is common in lower cost or pilot studies to have small samples for which large sample theory may not be valid. Specifically, we present the application of the derived exact CI, valid even in small samples, to the case of identifying the diagnostic state of renal functioning in patients post kidney transplant. 

This paper is organized as follows. Section \ref{sec:linearcombo} introduces the notation used to denote linear combinations of multinomial probabilities, derives the exact confidence interval, and outlines the computational implementation of the interval. In Section \ref{sec:sim}, a simulation study is used to demonstrate the coverage performance of the derived confidence interval under varying scenarios such as contrasts of multinomial probabilities and Bayes cost, and compare the new confidence interval method to existing methods. The simulation results show the proposed method maintains a minimum coverage of $1-\alpha$, unlike the large sample methods. Additionally, an average coverage adjusted variation on the developed interval, \textcolor{black}{useful when interval length is prioritized over exact confidence}, is presented. Section \ref{sec:can} demonstrates the use of the derived confidence interval by comparing multiple classifiers in the renal failure diagnostic problem with the Bayes cost metric. The simulation results and the application to detect diagnostic states after kidney transplant demonstrate the merits of this flexible interval. In Section \ref{sec:conc}, a final discussion is presented.

\section{Exact Interval for the Linear Combination of Multinomial Probabilities}
\label{sec:linearcombo}

\subsection{Definintion}
Let there be $K$ independent multinomial experiments, indexed on $%
k=1,\dots,K $. For the $k^{th}$ experiment there is a vector of random
variables $\mathbf{X_{k}}=(X_{1,k},\dots,X_{M_{k},k})$ where $\mathbf{X_{k}}$ is distributed multinomial, $\sim MN(\mathbf{p_{k}},n_{k})$, $M_{k}$ is the number of outcomes for
this experiment (indexed on $m_{k}=1,\dots,M_{k}$) \textcolor{black}{and $n_{k}$ is the sample size for the $k$th experiment}. Define a linear
combination of all multinomial probabilities from the $K$ independent
experiments as 
\begin{equation}\label{leq}\nonumber
L=\mathbf{p}^{\prime }\mathbf{w}
\end{equation}
where the vector \textcolor{black}{ $\mathbf{p}\in\mathcal{S}=\{\mathbf{p}=(\mathbf{p_{1}},\dots, \mathbf{%
p_{K}}): \mathbf{p_{k}}=(p_{1,k},\dots,p_{M_{k},k}), p_{m_{k},k}\ge
0$ \text{ and} $\sum_{m_{k}=1}^{M_{k}}p_{m_{k},k}=1, \forall  k\}$}, $%
\mathbf{w}$ is a vector of the constant multipliers (weights) in $\mathbb{R}$ to be
placed on each multinomial probability ($\mathbf{w}=(\mathbf{w_{1}},\dots, 
\mathbf{w_{K}})$, where $\mathbf{w_{k}}=(w_{1,k},\dots,w_{M_{k},k})$)\textcolor{black}{, and $\mathcal{L}=\{L=\mathbf{p}'\mathbf{w}: \mathbf{p}\in \mathcal{S}\}$}.

Define a $M_{k}\times 1$ vector $\mathbf{n_{k}}=(n_{k}^{-1},\dots
,n_{k}^{-1})$ for each multinomial experiment. The statistic used to
estimate $L$ is $\widehat{L}=Y=\left( \mathbf{X}\circ \mathbf{n}\right) ^{\prime }\mathbf{w}$ where $\circ $ represents the Hadamard product, and $\mathbf{n}$ and $\mathbf{X}$ are vectors such that $\mathbf{n}=(\mathbf{n_{1}}%
,\dots ,\mathbf{n_{K}})$, and $\mathbf{X}=(\mathbf{X_{1}},\dots ,\mathbf{%
X_{K}})$. Let $\mathcal{B}$ denote the joint multinomial sample space such
that $\mathcal{B}=\{\mathbf{x}=(\mathbf{x_{1}},\dots ,\mathbf{x_{K}}):%
\mathbf{x_{k}}=(x_{1,k},\dots ,x_{M_{k},k}),x_{m_{k},k}\in \mathbb{Z}%
^{+},\sum_{m_{k}=1}^{M_{k}}x_{m_{k},k}=n_{k},\forall k\}$. The observed
sample space for $Y$ consists of all the possible values of $\left( \mathbf{x%
}\circ \mathbf{n}\right) ^{\prime }\mathbf{w}$ that result from $\mathbf{x}%
\in \mathcal{B}$ and is denoted $\mathcal{Y}=\{y=\left( \mathbf{x}\circ 
\mathbf{n}\right) ^{\prime }\mathbf{w}:\mathbf{x}\in \mathcal{B}\}$.

\subsection{Exact Confidence Bounds}
\textcolor{black}{ \citet{WangFid} provides a simple and useful description of a $1 - \alpha$ fiducial interval on $\theta$ for an observed $Y=y$ as ``the answer to the question `What would be the possible values of $\theta$ which gave such a value $Y = y$ at the specified level $1 - \alpha$?'" Fiducial inference was first introduced by Fisher in his 1930 paper, ``Inverse Probability" \citep{Fisher1}.  However, in this and subsequent papers, Fisher did not fully develop his fiducial theory \citep{Fisher2, Pedersen1978, Hannig2009}. Due to this lack of development, controversy over the use of fiducial inference exists  \citep[see for example,][]{Pedersen1978, Zabell, Efron}. However, as \cite{Krish} note, objections to the fiducial approach are mainly philosophical, revolving around the under-developed theory and the inability to extend the fiducial approach universally, such as for multi-parameter distributions \citep{Pedersen1978, Zabell}. Despite the philosophical debate, the fiducial approach has practical and useful statistical inference applications often resulting in desirable frequentist properties, as is the case for linear combinations of multinomial parameters. }

Generally, a $(1-\alpha)100\%$ fiducial interval for a parameter $\theta$ derived from an observed statistic $Y=t(X_{1},\dots,X_{n})$ is found as the solution for $\theta_{L}$ and $\theta_{U}$ in the following two equations~\citep{WangFid}:
\begin{equation}
\label{fidl}
P(Y \ge y \mid \theta_{L}) = \alpha / 2
\end{equation}
\begin{equation}
\label{fidu}
P(Y \le y \mid \theta_{U}) = \alpha / 2.
\end{equation}

The fiducial percentiles of a one-dimensional parameter, found in Equations \ref{fidl} and \ref{fidu}, give an exact interval if the requirements for the fiducial argument  are met \citep{Pedersen1978, Zabell}. While Fisher never explicitly defined these requirements, other authors have derived them from Fisher's works \citep[see for example][]{Pedersen1978, Zabell}. 

Because our application involves projecting from a multi-dimensional to one-dimensional space, we cannot directly apply the one-dimensional logic of the fiducial argument. Instead, we carefully define the projection and resulting fiducial percentiles to ensure the exactness of the derived interval while meeting the requirements of the fiducial approach outlined in \cite{Pedersen1978}. 


To meet these requirements, we determine the solutions to Equations \ref{fidl} and \ref{fidu} for our parameter $L$ given an observed $y \in \mathcal{Y}$ as two fiducial percentiles for $L$, uniquely defined for the lower and upper bounds. First, we define
\textcolor{black}{
\begin{equation}\label{pdf1}
F_{Y}(y\mid \mathbf{p}) = P(Y \le y \mid \mathbf{p}) = \sum_{t \le y}\sum_{\substack{ \mathbf{x}\in 
\mathcal{B}  \\ \left(\mathbf{x}\circ \mathbf{n}\right) ^{\prime }\mathbf{w}=t}}f_{\mathbf{X}}(\mathbf{x}\mid \mathbf{p}).
\end{equation}}
\textcolor{black}{Equation \ref{pdf1} allows us to define the quantities we are interested in,  which are:}
\textcolor{black}{
\begin{equation}\label{CDF2}
F_{Y, LB}(y \mid L) = P(Y \ge y \mid L) =  1 - \inf_{\mathbf{p}: \mathbf{p}%
^{\prime }\mathbf{w} \le L}\left\{F_{Y}(y^{*}\mid \mathbf{p})\right\}
\end{equation}
and
\begin{equation}\label{CDF1}
F_{Y, UB}(y \mid L) = P(Y \le y \mid L) = \sup_{\mathbf{p}: \mathbf{p}%
^{\prime }\mathbf{w} \ge L}\left\{F_{Y}(y\mid \mathbf{p})\right\}
\end{equation}}
\textcolor{black}{where $y^{*}$ is the ordered value of $y \in \mathcal{Y}$ directly less than $y$. Also, $f_{\mathbf{X}}(\mathbf{x}\mid \mathbf{p})$ is the joint multinomial probability mass function for the underlying multinomial experiments generating $y$.  In Equations \ref{CDF2} and \ref{CDF1}, the infimum and supremum over \textcolor{black}{the subset a $\mathcal{S}$ where $\mathbf{p} \in \{\mathbf{p}: \mathbf{p}'\mathbf{w} \le L \}$ or  $\mathbf{p} \in \{\mathbf{p}: \mathbf{p}'\mathbf{w} \ge L \}$, for the lower and upper bound respectively,}  is required due to the projection from a multi-dimensional space into a one-dimensional space. \textcolor{black}{ By using the supremum and infimum, a unique solution to Equation \ref{CDF2} and \ref{CDF1} is found for any $L$, as for each $L$ the subset of $\mathcal{S}$ evaluated may be infinite but is bounded.} This approach ensures the distribution of $Y$ (\textcolor{black}{in our application, defined distinctly for the lower and upper bound}) depends entirely and only on $L$, and a solution can be found for any $L$ and is unique, a requirement for deriving fiducial percentiles. To demonstrate why this is true, we consider an example in Appendix 1.  Additionally, the choice of infimum or supremum \textcolor{black}{conservatively} ensures the exact coverage of the derived interval.}

\textcolor{black}{Then, the fiducial percentiles for the lower and upper bounds are:
\begin{equation}\label{LL}
L_{L} = L_{LB, \alpha / 2 }(y) =\inf_{L} \left\{ L \in\mathcal{L} \text{ such that } F_{Y,LB}(y \mid L) 
= \alpha / 2 \right\}
\end{equation}
\begin{equation}\label{LU}
 L_{U}=L_{UB, \alpha / 2 }(y) =\sup_{L} \left\{ L \in\mathcal{L} \text{ such that } F_{Y,UB}(y \mid L)
= \alpha / 2 \right\}.
\end{equation}}
\textcolor{black}{In Equations \ref{LL} and \ref{LU}, we include the infimum or supremum on the set of possible solutions due to our projection from a multi-dimensional to a one-dimensional space. This approach ensures unique solutions for $L_{L}$ and $L_{U}$ for any $y \in \mathcal{Y}$ and $\alpha \in (0,1)$, an additional requirement for $L_{U}$ and $L_{L}$ to be fiducial percentiles of $L$. Achieving uniqueness by taking the supremum or infimum of the \textcolor{black}{fiducial percentile} set is allowable under the fiducial approach \citep{Pedersen1978}, and the choice of supremum for $L_{U}$ and the infimum for $L_{L}$ \textcolor{black}{conservatively} ensures the exactness of the interval. }

 \textcolor{black}{$L_{LB, \alpha / 2 }(y)$ is the $L$ }for which $y$ belongs to the $\alpha / 2$ percentiles of $F_{Y,LB}(\cdot \mid L)$, where the $\alpha/2$ percentiles of $F_{Y,LB}(\cdot \mid L)$ are denoted $y_{LB, \alpha/2}(L)$ and defined 

\begin{equation}\label{yalphaLB}
y_{LB, \alpha/2}(L) = \min_{y} \left\{ y \in \mathcal{Y}  \text{ such that } F_{Y,LB}(y \mid L)
\le \alpha/2 \right\}.
\end{equation}
Additionally, $L_{UB, \alpha / 2 }(y)$ is the \textcolor{black}{$L$} for which $y$ belongs to the $\alpha / 2$ percentiles of $F_{Y, UB}(\cdot \mid L))$, where the $\alpha/2$ percentiles of $F_{Y, UB}(\cdot \mid L)$ are denoted $y_{UB, \alpha/2}(L)$ and defined 
\begin{equation}\label{yalphaUB}
y_{UB, \alpha/2}(L) = \max_{y} \left\{ y \in \mathcal{Y}  \text{ such that } F_{Y, UB}(y \mid L)
\le \alpha/2 \right\}.
\end{equation}

Meeting the final requirement for $L_{U}$ and $L_{L}$ to be fiducial percentiles of $L$ is the fact that $y_{UB, \alpha/2}(L)$ and $y_{LB, \alpha/2}(L)$ are non-decreasing in $L$ (Proof in Appendix 2). Equations \ref{yalphaLB} and \ref{yalphaUB} use a minimization and maximization, respectively,  on the $y_{\alpha}(L)$ set instead of infimum and supremum because $\mathcal{Y}$ is finite. Additionally, Equations \ref{yalphaUB} and \ref{yalphaLB}  use $\le$ instead of an equality because $\mathcal{Y}$ is discrete.  $L_{U}$ and $L_{L}$ provide coverage greater or equal than $ 1-\alpha$ for all $L$. Proof is provided in Appendix 2.

\subsection{Computational Implementation}
\label{sec:computation}
\color{black}

We present an algorithm for computing the bounds on $L$ for a given $\widehat{L}$ defined in Equations \ref{LL} and \ref{LU}. This algorithm requires two processes and two simplifying assumptions. The assumptions are reasonable with respect to exact confidence due to the conservative nature of fiducial intervals for probabilities noted by other authors \citep[see, for example,][]{AgrCoull, Thulin2012}, and demonstrated with simulation in Section \ref{sec:sim}.  
\begin{myassump}
$y_{UB, \alpha/2}(L)$ and $y_{LB, \alpha/2}(L)$ are still non-decreasing in $L$ when replacing the inequalities in Equations \ref{CDF2} and \ref{CDF1} with equalities as follows:
\begin{align}
F_{Y, LB}(y \mid L) &= 1 - \inf_{\mathbf{p}: \mathbf{p}^{\prime }\mathbf{w} = L}\left\{F_{Y}(y^{*}\mid \mathbf{p})\right\} \ \text{ and,} \label{assump1}\\
F_{Y, UB}(y \mid L) &= \sup_{\mathbf{p}: \mathbf{p}^{\prime }\mathbf{w} = L}\left\{F_{Y}(y\mid \mathbf{p})\right\}. \label{assump2}
\end{align}
\end{myassump}
\begin{myassump}
A stochastic optimizer provides an adequately close approximate of $F_{Y, LB}(y \mid L)$ and $F_{Y, UB}(y \mid L)$, as defined in Equations \ref{assump1} and \ref{assump2}.
\end{myassump}
\noindent
From our investigations through simulation, these assumptions apear to be valid. With these two assumptions, the bounds are found stochastically using the procedures as outlined below.\\
\noindent
\textbf{Process 1}:  Finds the CDF of $\widehat{L}$ = $Y$ (i.e., $F_Y(y \mid \mathbf{p})$) for a given $\mathbf{p}$. First, find all possible values of $Y$ (on an evenly spaced grid, some values on the grid will not be possible values of $Y$). Then, calculate the Fourier transform (FT) of each independent multinomial trial (which comprises the vector of multinomials) multiplied by the appropriate weight. 
These FTs are multiplied together, which yields the FT for $Y$. The PMF for $Y$ is then obtained using the inverse discrete Fourier transform (iDFT). The inversion of $Y$'s FT to the PMF can be obtained precisely (neglecting numerical computing error) using the methods found in \cite{warr2014numerical}. Finally, cumulatively sum the PMF to obtain the CDF.

\noindent
\textbf{Process 2}: Finds $F_{Y, LB}(y \mid L)$ and $F_{Y, UB}(y \mid L)$ in Equations \ref{assump1} and \ref{assump2} for a given $L$ and requires the use of Process 1.
In this process, the space of possible values for the vector $\mathbf{p}$ (constrained such that $\mathbf{p}’  \mathbf{w}=L$) is explored randomly for a fixed number of iterations (say $n_r$). This will result in an initial approximated optimum for Equation \ref{assump1} or \ref{assump2}. Then, the variance is decreased as the current estimate of the approximated optimum is used to search for better solutions that are nearby (for $n_s$ iterations).  \textcolor{black}{This approach is similar in nature to simulated annealing, a process in which the temperature of a system is slowly decreased to allow the system to settle into a state of minimum energy as summarized in} \cite{brooks1995optimization}.  Thus, a vector $\mathbf{p}$ is obtained that results in values close to the infimum or supremum in \ref{assump1} and \ref{assump2}, respectively. 

In other words, for a given $L$, ``Randomly'' sample $\mathbf{p}$ (constraining $\mathbf{p}$ such that $\mathbf{p}'\mathbf{w}=L$) $n_r$ number of times, \textcolor{black}{where $n_{r}$ is a tuning parameter}. Note, the sampling scheme on $\mathbf{p}$ does not weight each eligible $\mathbf{p}$ equally, but this does not create an issue when optimizing. Of the $n_r$ samples of $\mathbf{p}$, Process 1 gives $F_{y}(y \mid \mathbf{p})$ and then we find the $\mathbf{p}$ such that $F_Y(y^*\mid \mathbf{p})$ is smallest as in Equation \ref{assump1} (for the lower bound), or $F_Y(y \mid \mathbf{p}$) is the largest as in Equation \ref{assump2} (for the upper bound). Using the previously found $\mathbf{p}$, stochastically perturb $\mathbf{p}$ and test if it produces a better value \textcolor{black}{(i.e., smaller or larger for the lower or upper bound, respectively)}, if so, use that value as the best $\mathbf{p}$ and perturb it to search for another value (do this $n_s$ number of times, decreasing the variance of the perturbation each iteration, \textcolor{black}{where $n_{s}$ is another tuning parameter.}). This provides the estimated values of $F_{Y, LB}(y \mid L)$ and $F_{Y, UB}(y \mid L)$ for a given $L$.  

\noindent
\textbf{Confidence Interval Bound Algorithm}: Step 1: Pick an $L$ from the possible valid values.  Step 2: For the given $L$, find $F_{Y,LB}(y\mid L)$ using Process 2. If $F_{Y,LB}(y\mid L)$ is sufficiently close to $\alpha/2$, set $L_L=L$ and move to the next step, otherwise return to Step 1. Step 3: Pick an $L$ from the possible valid values.  Step 4: For the given $L$, find $F_{Y,UB}(y\mid L)$ using Process 2. If $F_{Y,UB}(y\mid L)$ is sufficiently close to $\alpha/2$, set $L_U=L$ and quit, otherwise return to Step 4.

Two additional points for implementing the confidence interval algorithm are discussed here briefly.  First, being ``sufficiently close to $\alpha/2$''  requires setting a desired numerical precision.  This is typically done in the context of the application.  In some applications a high degree of precision is required while others may only need one or two decimals of precision.  Second, picking a possible value of $L$ is an important aspect of the algorithm's efficiency.   In our implementation we use a root-finding method on the equations $f(L) = F_{Y,LB}(y\mid L)-\alpha/2$ for the lower bound and $g(L) = F_{Y,UB}(y\mid L)-\alpha/2$ for the upper bound.  The roots to these two equations are the approximated bounds of $L$ for a given $\widehat{L}$.  The root-finding method makes the decisions of finding new values of $L$ and attempts to efficiently find the value of $L$ that satisfies our requirements.  The root-finding method we employ is the \textit{uniroot} function in the \textit{utils} package of R \citep{R}.

The computation times for this algorithm vary depending on a few factors.  The primary factor being how many possible values of $\widehat{L}$ exist \textcolor{black}{(i.e., the cardinality of $\mathcal{Y}$) for the given $\mathbf{w}$ and $\mathbf{n}$.}  The others are the numbers $n_r$ and $n_s$, which can be varied according to preference.  We have found that relatively small values, $n_r=20$ and $n_s=20$, produce reasonable results.  The computation times to find confidence bounds for one $\widehat{L}$ in the simulations for Section \ref{sec:sim} range from $\approx$ 2 - 8 seconds depending on the complexity of the scenario.  However, the computation time increases dramatically in the application in Section \ref{sec:can}, with computation times around 360 seconds.  All times are with $n_r=n_s=20$.

This algorithm underestimates the upper bound and overestimate the lower bound.  However, given the conservative nature of the \textcolor{black}{theoretical interval bounds in Equations \ref{LL} and \ref{LU}}, even a close approximation should produce ``exact'' intervals (as demonstrated in the simulations).  In practice, we recommend finding the bounds a few times to gain confidence that the solutions are roughly equivalent.
\section{Simulation Study}
\label{sec:sim}
\subsection{Scenarios}
The performance of the derived exact confidence interval around linear combinations of multinomial probabilities, as measured by \textcolor{black}{coverage probability and interval length, is demonstrated with simulation.} \textcolor{black}{Four} different linear combination scenarios are considered across four small-sample sizes ($n_{j}=5,10, 15$ and $20$) to demonstrate a range of performance.  The first scenario, denoted Scenario A, consists of three multinomial experiments ($K=3$) with three outcomes each ($M_{k}=3$). This scenario reflects a possible Bayes cost metric from a diagnostic test with three classes, where $\mathbf{w}=(w_{1,1},w_{2,1},w_{3,1},w_{1,2},w_{2,2},w_{3,2},w_{1,3},w_{2,3},w_{3,3})=(0,1,1,2,0,3,5,3,0)$.  The second scenario, denoted Scenario B, consists of two multinomial experiments ($K=2$) with four outcomes each ($M_{k}=4$) and $\mathbf{w}=(1,2,3,0,1,1,2,0)$. Scenario C consists of a simple contrast for two multinomial experiments ($K=2$), each with two outcomes ($M_{k}$=2) where $\mathbf{w}=(1,0,-1,0)$. \textcolor{black}{Finally, Scenario D demonstrates a more complex contrast of two multinomial experiments ($K=2$) with differing number of outcomes. The first experiment is a multinomial with three outcomes ($M_{1}=3$) with weights contrasting the first class and the sum of the second and third classes. The second experiment is a multinomial with four outcomes ($M_{2}=4$) with weights contrasting the first class with the remaining three classes.  The concatenated weight vector for the first and second experiment is $\mathbf{w}$ = (4, -2, -2, 4, -1, -1, -2).  The interesting note here is that for both experiments, the first class contrasts the remaining classes but the number of remaining classes differ between the experiments.  Such scenarios are feasible, for example, when a legacy technology (the first experiment) is combined with newer technology (the second experiment) that is able to differentiate  classes which previously could not be separated. This final scenario is chosen to illustrate the flexibility of the method, and therefore only a sample size of $n=10$ is simulated.} A significance level of $\alpha=0.05$ is assumed for all scenarios \textcolor{black}{and intervals are calculated as outlined in Section \ref{sec:computation} using $n_r=n_s=20$.}

\subsection{Simulation Approach}\label{simapproach}
\textcolor{black}{For any $L$, interval coverage probability is measured as \citep{Rubin1986}:
\begin{equation}\label{ConfCov}
c(L)=\sum_{y \in \mathcal{Y}}I(L\in[L_{L},L_{U}])f_{Y}(y \mid L).
 \end{equation}
Confidence coefficient is the infimum of coverage probabilities across the parameter space, 
\begin{equation}\label{ConfCoeff}
\text{Confidence Coefficient}=\inf_{\mathbf{p}} \left \{ c(L) \right\}.
\end{equation}}
\textcolor{black}{For all scenarios, coverage probability is estimated across the range of $L$, using 1,000 unique and uniformly distributed $L$ values. Then for each $L$, we randomly draw 1,000 $\mathbf{p}$, such that   $\mathbf{p}'\mathbf{w}=L$. For our exact CI, the coverage probability at each $\mathbf{p}$ is calculated directly using Equation \ref{ConfCov}. For the large sample methods, the coverage probability for each $\mathbf{p}$ is estimated with Monte Carlo simulation using 500 draws of $\mathbf{x}$ for each $\mathbf{p}$, as described in \citet{Rubin1986}. Then, the average coverage across the 1,000 $\mathbf{p}$ estimates the coverage probability for each $L$. Finally, the confidence coefficient, Equation \ref{ConfCoeff}, is estimated as the minimum coverage across all $\mathbf{p}$ (1,000 $\mathbf{p}$ $\times$ 1,000 $L$ = 1,000,000 $\mathbf{p}$)  for each scenario.  }

\subsection{Simulation Results}

\begin{figure}[p]
\centering
    \includegraphics[width=.9\linewidth]{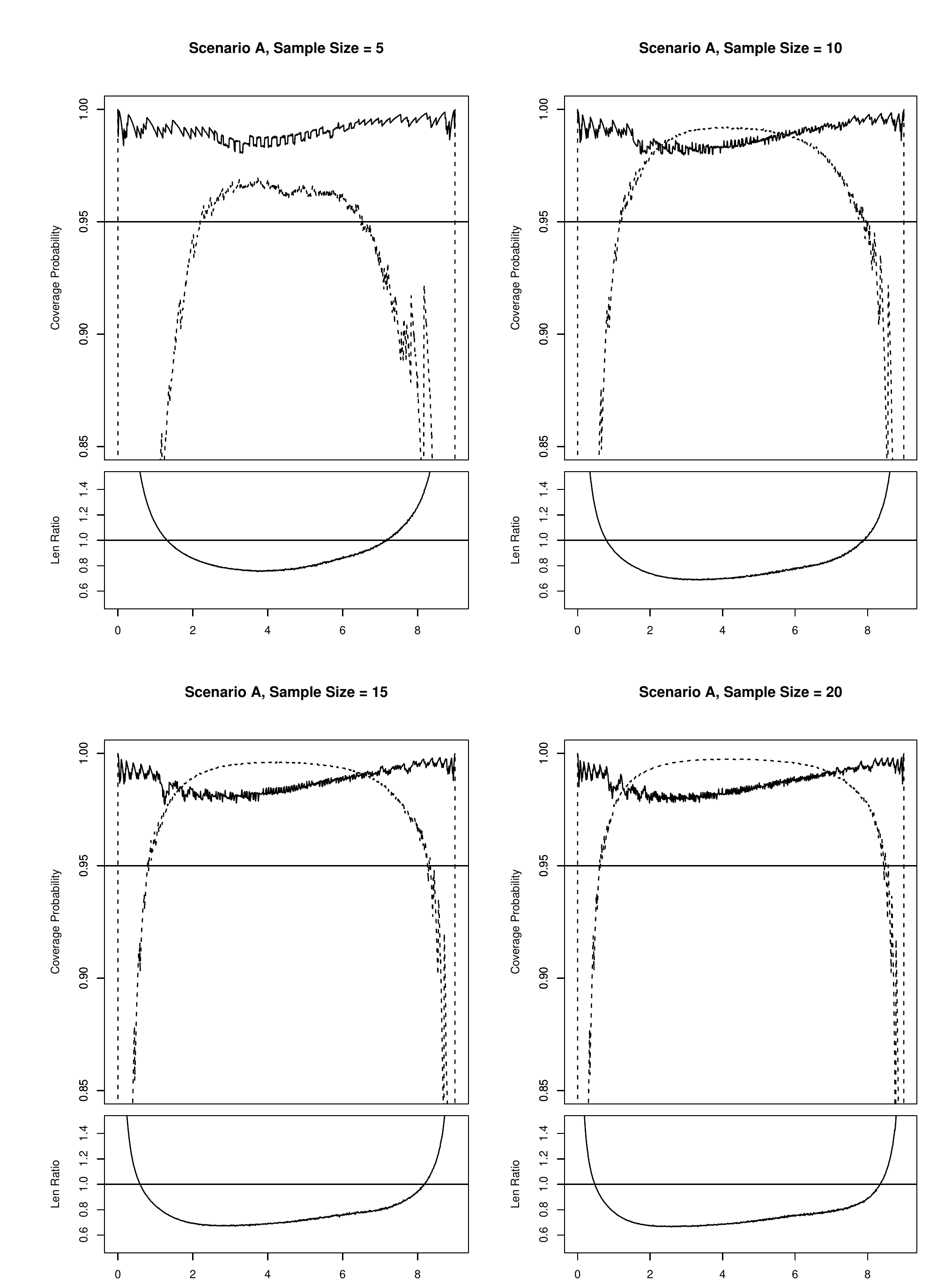}
        \caption{Simulation results for Scenario A. The solid line in the top chart in each quadrant is the fiducial interval coverage probability. The solid line in the bottom chart in each quadrant is the ratio of the fiducial interval length to the Gold interval length. The dashed line is the Gold interval coverage probability.  Reference lines at 0.95 coverage and a length ratio of 1 are also graphed.}
    \label{SimA}
\end{figure}

\begin{figure}[p]
\centering
    \includegraphics[width=.9\linewidth]{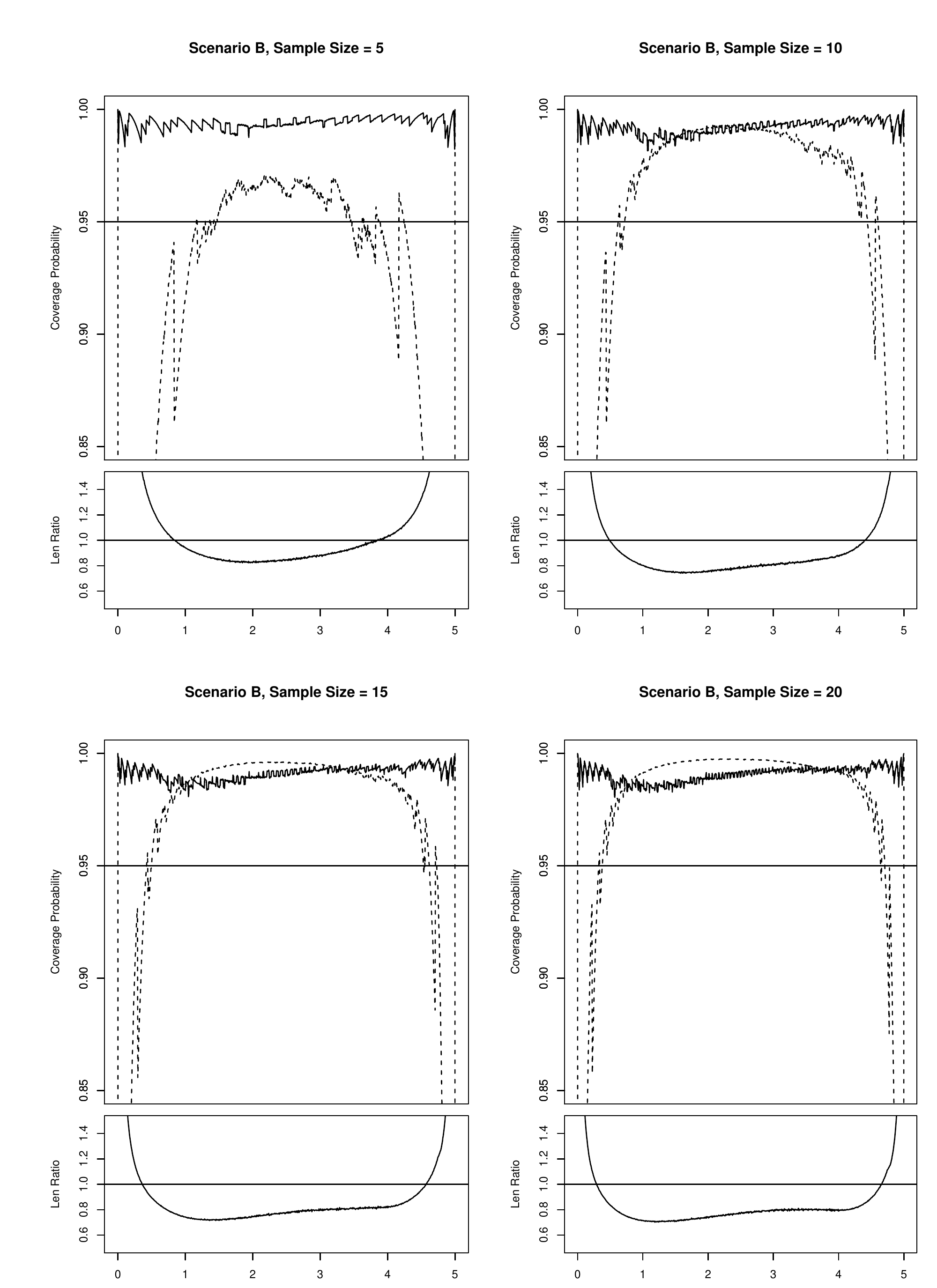}
        \caption{Simulation results for Scenario B. The solid line in the top chart in each quadrant is the fiducial interval coverage probability. The solid line in the bottom chart in each quadrant is the ratio of the fiducial interval length to the Gold interval length. The dashed line is the Gold interval coverage probability. Reference lines at 0.95 coverage and a length ratio of 1 are also graphed.}
    \label{SimB}
\end{figure}

\begin{figure}[p]
\centering
    \includegraphics[width=.9\linewidth]{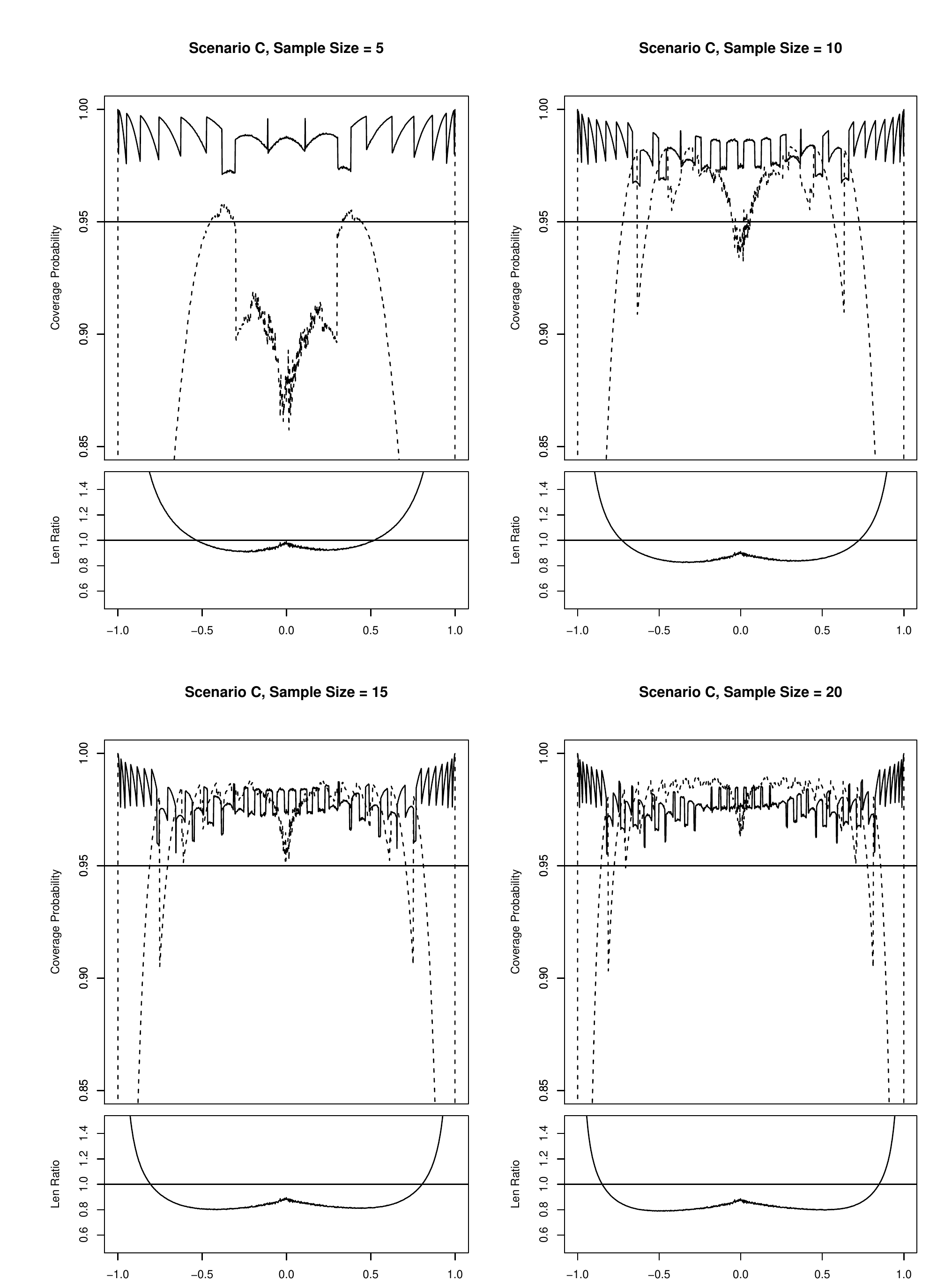}
        \caption{Simulation results for Scenario C. The solid line in the top chart in each quadrant is the fiducial interval coverage probability. The solid line in the bottom chart in each quadrant is the ratio of the fiducial interval length to the Goodman interval length. The dashed line is the Goodman interval coverage probability. Reference lines at 0.95 coverage and a length ratio of 1 are also graphed.}
    \label{SimC}
\end{figure}

For Scenarios A and B, the coverage probability and length of the derived confidence interval is compared to the large sample method for confidence intervals around linear combinations of multinomial probabilities developed in \citet{Gold1963}.  The performance of the exact interval around contrasts, Scenario C, is compared to the method presented in \citet{Goodman1964}, which adjusts the chi-square degrees of freedom from \citet{Gold1963} for use specifically with contrasts. 
All \textcolor{black}{four} scenarios demonstrate the exact nature of the interval presented in this paper, with coverage \textcolor{black}{well} above 95\% \textcolor{black}{reflecting conservative average coverage across the range of $L$ values}. The simulation results for Scenarios A through C are presented in Figures \ref{SimA} to \ref{SimC}.

As expected, the large sample methods do not maintain coverage above 95\%, \textcolor{black}{especially in the smaller sample examples. Even in larger samples, the coverage for the large sample methods drops below 95\% for various values of $L$, including the tails. When the large sample method does achieve coverage $\ge$ 95\%, the large sample coverage is often as conservative as the exact method with the exact method consistently having lengths shorter than the large sample method. Evaluating the minimum coverage across all $\mathbf{p}$ with the derived exact confidence interval gave estimated confidence coefficients for samples sizes of 5, 10, 15, and 20 of 0.9511, 0.9510, 0.9496, and 0.9496 for Scenario A, 0.9468, 0.9492, 0.9490, and 0.9493 for Scenario B, and 0.9611, 0.9580, 0.9506, and 0.9518 for Scenario C.}

\textcolor{black}{Finally, the large sample methods available are not designed to handle Scenario D's experiments with differing numbers of outcomes, but the complexity of this scenario does not create challenges for our developed method. Because no comparison is made, results are presented as average coverage and confidence coefficient without a figure. The exact interval still ensures coverage $\ge 95\%$ across all $L$ values. Specifically, for Scenario D, the exact interval has an average coverage across $L$ of 0.9778 and a confidence coefficient of 0.9492.} Overall, all simulation results presented demonstrate the very desirable characteristics of the developed exact interval for the linear combination of multinomial probabilities.
 
\subsection{Coverage Adjusted Bounds}
\label{sec:covadj}

Like the Clopper-Pearson interval for a binomial proportion, the confidence interval for the sum of multinomial proportions presented in this paper ensures minimum coverage of at least $1-\alpha$ across all $L$. As demonstrated in the simulation results, such coverage is not accomplished with large sample methods. While achieving the minimum coverage across $L$ is required of an exact interval, this often results in conservative coverage probabilities, often greater than the desired level of $1-\alpha$. 
 
As an alternative, some authors have suggested that average coverage may be more useful than minimum coverage when constructing intervals for  proportions~\citep{AgrCoull, BrownCai2001, Newcombe2011}. Therefore,~\citet{Thulin2012} proposed an average coverage adjusted confidence interval for the Clopper-Pearson confidence interval. If less conservative bounds are of interest, such an adjustment may also be extended for the exact interval on the linear combination of multinconfidence interval omials presented in this paper. To derive the adjusted interval bounds, an adjusted alpha, $\alpha' \ge \alpha$, is chosen such that the average coverage, instead of the minimum coverage, equals $1-\alpha$. The required $\alpha'$ is found consistent with the method presented in \citet{Thulin2012} such that:

\begin{equation}\nonumber
C(\alpha') = \int_{\mathcal{L}}\sum_{y\in \mathcal{Y}} I (L \in [L'_{L}, L'_{U}])f_{Y}(y \mid L)f(L)dL=1-\alpha.
\end{equation}
Here, $I (L \in [L'_{L}, L'_{U}])$ is an indicator function resulting in 1 if $L$ is in the fiducial interval constructed with $\alpha'$ ($[L'_{L}, L'_{U}]$,  defined in Equations \ref{LL} and \ref{LU}) and 0 otherwise. Additionally, $f(L)$ allows for a distribution on the values of $L$. In this work, all $L \in \mathcal{L}$ are assumed equally likely, fixing $f(L)$ \textcolor{black}{$\propto$} 1.

\textcolor{black}{Finding the adjusted coverage fiducial bounds requires significant additional computation.  First, the exact bounds should be found for all possible $\widehat{L}$.  Then, the coverage should be estimated for all $L$ using the procedure in Section \ref{simapproach}, requiring the confidence intervals for each $\widehat{L}$. From this estimate of coverage, the average coverage can easily be calculated.  When $\alpha=\alpha'$, the average coverage is expected to be higher than $1-\alpha$.  Choosing several $\alpha'$ greater than $\alpha$ and repeating this procedure, an $\alpha'$ that produces an approximate average coverage of $1-\alpha$ is found.}

\begin{figure}[]
\centering
    \includegraphics[width=.4\linewidth]{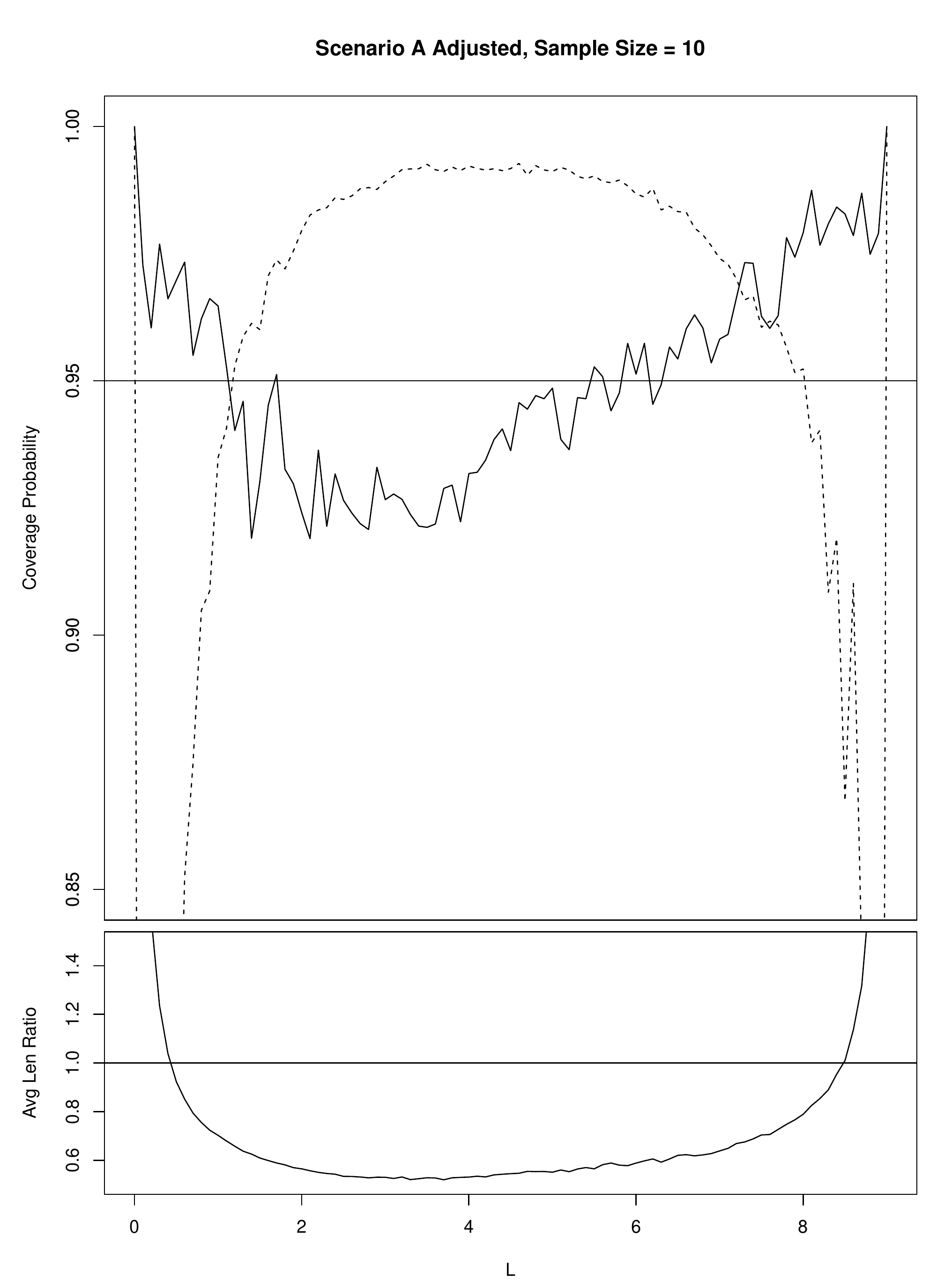}
        \caption{Average Coverage Adjusted Interval Simulation results for Scenario A, n=10. Top chart coverage probability: average coverage adjusted fiducial interval (dark solid line), Gold interval (dashed line). Bottom chart length ratios: average coverage adjusted fiducial interval to Gold interval (solid line), average coverage adjusted fiducial interval to Gold (dashed line).  }
    \label{SimAAdj}
\end{figure}

To demonstrate this adjustment, the average coverage interval for Scenario A with n = 10 is constructed. The adjusted $\alpha'$ found to achieve an average coverage of approximately 0.95 for this example is \textcolor{black}{$\alpha' =0.138$.} The simulation was re-run with these adjusted bounds on $L$ and plotted with the \textcolor{black}{Gold interval} coverage in Figure \ref{SimAAdj}. First, it is clear the coverage probability of this interval does not ensure $1-\alpha$ across all possible $L$ as the original exact confidence interval does. However, the adjusted interval still has better coverage probability overall than the large sample method across the range of $L$. Additionally, the adjusted bounds have length approximately \textcolor{black}{76\% }the length of the original confidence interval \textcolor{black}{and on average, 66\% the length of the Gold interval (excluding the end points where the ratio is infinite)}. Therefore, the adjusted confidence interval provides even shorter lengths than both the original exact interval and the large sample methods while still achieving better coverage than the large sample method overall (for Scenario A, n=10, the Gold method had a simulation estimated \textcolor{black}{93\% }average coverage and confidence coefficient of zero whereas the adjusted interval has an \textcolor{black}{estimated average coverage of 95\% and }confidence coefficient of \textcolor{black}{85\%}). 

\section{Application to Medical Diagnostics}
\label{sec:can}

To demonstrate the use of the confidence interval presented in this paper we examine the medical classification of the diagnostic state of renal functioning in patients post transplant. Chronic allograph nephropathy is one of the primary conditions leading to renal transplant failure after kidney transplant \citep{CANpaul, CAN1}.  Its progression from restored normal kidney function as a result of transplant, however, is still not well understood.  Biopsy remains the means to fully diagnose chronic allograph nephropathy, though biopsy is an expensive, invasive procedure which may result in a negative finding.  Instead, the inflammatory response of tissue damage which is associated with chronic allograph nephropathy may be a non-invasive mechanism for diagnosis.  Proinflammatory cytokine markers, such as the transforming growth factor-$\beta$1,  have been the focus of research on the early indication of allograft loss~\citep{CANpaul}. ~\citet{CAN1} conducted a study to determine if the cytokines from the gene panel mRNAs in urine could be used to classify patients six months post-transplant as either having normal kidney function, normal kidney function with proturnia, or as having chronic allograph nephropathy. This study found the biomarkers transforming growth factor-$\beta$1 (TGF-$\beta$1), angiotensinogen (AGT), and epidermal growth factor receptor (EGFR) could be useful as early predictors of allograft function~\citep{CAN1}.  

Using these three biomarkers, three different classification methods were considered for distinguishing between the three kidney functioning classes.  The three classification methods considered were: a simple recursive partitioning algorithm; traditional parametric multinomial regression; and a bayesian additive regression tree. The simple recursive partitioning method assigns an individual to the chronic allograph nephropathy class if TGF-$\beta1 > \theta_{1}$, normal kidney function with proturnia  if TGF-$\beta$1 $\ge \theta_{1}$ and AGT $ > \theta_{2}$, and normal kidney function otherwise.  The optimal values for $\theta_{1}$ and $\theta_{2}$ used in the algorithm are those that minimized misclassifications.  The simple recursive partitioning method presented only uses two biomarkers as this provided equivalent classification when compared to the use of three biomarkers.  For the multinomial regression, parameters were solved traditionally using maximum likelihood based upon the linear combination of the three biomarkers, and the bayesian additive regression tree algorithm used the default settings in R \citep{R, RBART}. In Figure \ref{CANTab2} solutions from each of these classifiers were consolidated into 3 by 3 contingency tables.

\begin{table}
\begin{tabular}{cccc}
&&Normal&\\
&Normal &kidney& Chronic \\ 
&kidney &function& allograph \\ 
&function &w/ proturina& nephropathy \\ 
\cline{2-4}
& \multicolumn{3}{c}{simple recursive partitioning}\\
\cline{2-4}
Normal kidney function  &26&1&5  \\
Normal kidney function with proturnia&5&9 & 4  \\
Chronic allograph nephropathy &1&2 &11 \\
\cline{2-4}
& \multicolumn{3}{c}{bayesian additive regression tree}\\
\cline{2-4}
Normal kidney function  &  29&1  & 2  \\
Normal kidney function with proturnia& 5 &  10& 3 \\
Chronic allograph nephropathy &2  &  2 & 10 \\
\cline{2-4}
& \multicolumn{3}{c}{multinomial regression}\\
\cline{2-4}
Normal kidney function  &30  &2  &0 \\
Normal kidney function with proturnia& 11 & 7 &0 \\
Chronic allograph nephropathy  & 2 & 8 &4\\
\end{tabular}
\caption{Classification results from three classifiers (rows = truth, columns = classification outcomes).}
\label{CANTab2}
\end{table}

This leads to ask, which of these three classifiers provides the most desirable classifier? Earlier work sought to answer this question, but without small-sample inference on linear combinations of multinomial probabilities, the Youden index, \textcolor{black}{i.e., the sum of the correct classification probabilities,} was used instead of Bayes cost,  \textcolor{black}{i.e., the sum of the weighted mis-classification probabilities} \citep{Batterton2}.  Bayes cost allows for a more detailed consideration of a diagnostic test \textcolor{black}{with three or more diagnostic outcomes}, and is therefore used in this paper for further exploration and refinement of this diagnostic problem \citep{RefWorks40, RefWorks17, RefWorks30, RefWorks41, Skaltsa2010, RefWorks6, RefWorks22, Batterton1}. 

Bayes cost is the weighted sum of the $K^{2}-K$ misclassification probabilities resulting from a $K$-class classification system.  
In general, Bayes cost is a special case of $L$, where each class has the same number of outcomes, $M_{k}=K$ indexed on $m=1,\dots,K$. Additionally, $\mathbf{w_{k}}=(c_{1,k}\times pr_{k},\dots,c_{K,k}\times pr_{k}):c_{m=k, k}=0$. Here, $c_{m,k}$ is the cost of misclassifying disease class $k$ as $m$ and $pr_{k}$ is the prevalence of the $k^{th}$ class. Additionally, Bayes cost only includes misclassification probabilities as there is no cost associated with correct classification,  resulting in the restriction that all $c_{m=k, k}=0$. For new diagnostic tests, the optimal operating thresholds for the test may be found as those that minimize Bayes cost. However, Bayes cost may be estimated from the classification results of any classifier, regardless of how or on which optimal criterion the classifier is derived.  \textcolor{black}{In this section, we use $BC$ in place of $L$ to reference the specific linear combination of multinomial probabilities employed. }


In the chronic allograph nephropathy application, we have a 3-class classification system (classes: chronic allograph nephropathy, normal kidney function with proturnia  and normal kidney function), with six potential misclassifications because there are two possible errors for each true class. For example, a person who has chronic allograph nephropathy may be incorrectly diagnosed as normal kidney function with proturnia  or normal kidney function. In the case of an individual who has chronic allograph nephropathy, depending on the prescribed interventions, financial cost, and risks to non-intervention, the diagnostic cost of misdiagnosing the individual as normal kidney function may be different from the diagnostic cost of diagnosing this individual as normal kidney function with proturnia .  Additionally, we have set the prevalence of the normal kidney function, normal kidney function with proturnia  and chronic allograph nephropathy classes as 0.50, 0.28, and 0.22 respectively, matching closely to those found in \citet{CANpaul} and \citet{Khan2014}.  Assuming false negatives are more costly than false positives, misclassification weights were set such that if a chronic allograph nephorpathy patient was misdiagnosed as normal kidney function $c=  45$ and if this patient was misdiagnosed as normal kidney function with proturnia $c= 14$. If a normal kidney function with proturnia patient was misdiagnosed as normal kidney function $c = 25$ and if this patient  was misdiagnosed with chronic allograph nephropathy $c=4$. Finally, if a normal kidney function patient was misdiagnosed with normal kidney function with proturnia $c=4$ and if this patient was misdiagnosed with chronic allograph nephropathy $c=4$.  The misclassification costs and prevalences produced a weight vector for Bayes cost, rounded to the nearest whole number. 

While large sample, parametric inference methods for Bayes cost and multinomial probabilities exist~\citep{Skaltsa2010, RefWorks6, RefWorks22, Batterton1}, the total sample size \textcolor{black}{of availbale data} from~\citet{CAN1} is 64 patients, divided among three classes. Specifically, class sample size is 14, 18 and 32 for chronic allograph nephropathy, normal kidney function with proturnia , and normal kidney function respectively.  The application of a large sample procedure may therefore not be valid; instead, the exact confidence interval for the linear combination of multinomial probabilities presented in this paper is very useful and perhaps more appropriate for this problem set. 

\begin{figure}[]
\centering
    \includegraphics[width=.5\linewidth]{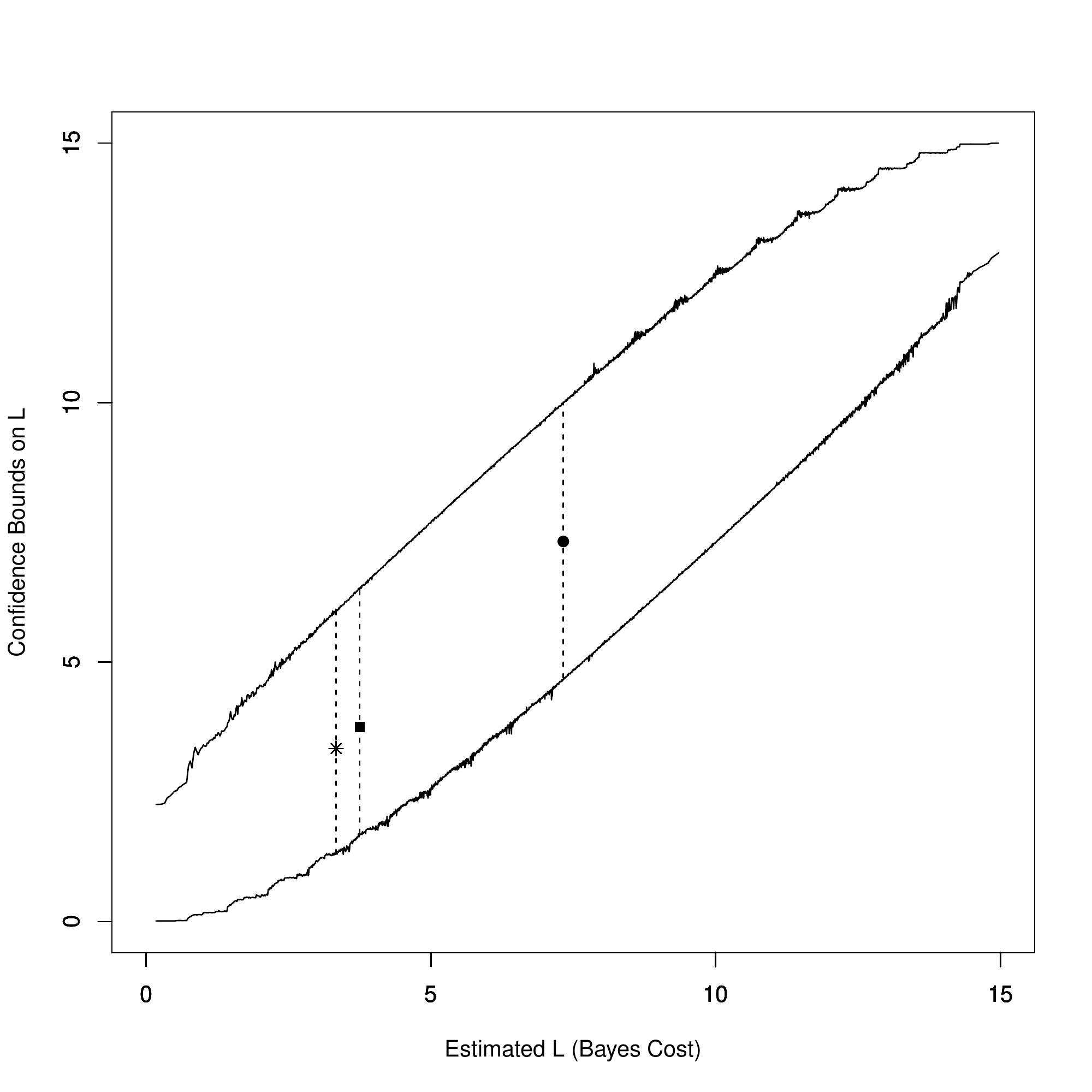}
        \caption{The exact confidence interval upper and lower bounds for the chronic allograph nephropathy classification problem plotted across all possible Bayes cost (solid lines) with the estimated Bayes cost for each classifier: simple recursive partitioning (star), bayesian additive regression tree (square), multinomial regression (circle). The upper and lower bound for each classifier is connected with a dashed line.}
    \label{CANCI}
\end{figure}
  
 Figure \ref{CANCI} provides the 95\% exact bounds on Bayes cost for all possible estimates of Bayes cost, $\widehat{BC}$, and the three classifiers' $\widehat{BC}$ and 95\% exact confidence intervals. Simple recursive partitioning and bayesian additive regression tree had similar overall unweighted classification rates (28.2\% and 23.4\% misclassifications, respectively),  \textcolor{black}{but their performance ordering changes when considering} their class-specific misclassifications. Specifically, simple recursive partitioning \textcolor{black}{does a better job} classifying chronic allograph nephropathy patients compared to the other two classifiers. Because of the heavier weighting placed on the cost of not detecting chronic allograph nephropathy, the simple recursive partitioning Bayes cost value is lowest, \textcolor{black}{3.336 (95\% CI: 1.307, 5.985)}. This distinction between simple recursive partitioning and bayesian additive regression tree is not possible without considering the class specific misclassifications. The bayesian additive regression tree had the second lowest $BC$ value, \textcolor{black}{3.751 (95\% CI: 1.698, 6.389)}, and multinomial regression the largest, \textcolor{black}{7.324 (95\% CI: 4.695, 10.011).} 
     
While the 95\% confidence intervals for the three classifiers overlap, the exact confidence intervals provide insight into the uncertainty in the point estimates with an appropriate confidence level despite the small sample size.  The exact confidence interval developed in this paper demonstrates to researchers that additional data may be required to reduce uncertainty and select the best classifier, or alternate classifiers may be explored. If a less conservative interval is desired, the average coverage adjusted fiducial interval may be used instead of the exact, providing narrower intervals with good coverage across $BC$. 
          
Diagnosing chronic allograph nephropathy post kidney transplant is an important medical problem for which inference on the linear combinations of multinomial probabilities proves useful. Developing a good diagnostic method that is not invasive requires exploration into the best combination of biomarkers and classification techniques while also considering the costs of class specific misclassifications. This was clearly demonstrated in our exploration of the chronic allograph nephropathy diagnostic problem, as an unweighted metric indicated a bayesian additive regression tree classifier performed best while the weighted metric, a linear combination of multinomial probabilities, indicated more desirable classification with the simple recursive partitioning classifier. The exact confidence intervals on these weighted measures allows for an understanding of the uncertainty in these estimates, demonstrating the superiority between the bayesian additive regression tree and the simple recursive partitioning is not clear. Finally, while this paper focuses on medical diagnostics, our exact confidence interval can be used to compare performance of any classification problem as long as the outcomes can be expressed as a linear combination of multinomial probabilities (e.g., Bayes cost, expected utility). 
    
\section{Discussion}
\label{sec:conc}

The solutions for the exact confidence interval bounds on $L$, derived in this paper with the fiducial approach, do not have a closed form solution. However, computational methods available today, 90 years after the first introduction of the fiducial approach, make its extension to linear combinations of multinomial probabilities feasible. Specifically, we solve the multinomial cumulative distribution function with a fast Fourier transform and find approximate optimal solutions by combining a numerical solver and stochastic optimizer. With these methods, we are able to implement a fairly complex optimization problem of finding the exact confidence bounds in a reasonable and practically useful timeframe. For example, for our simulation scenarios with sample sizes of 20 the solution for a single observed $\widehat{L}$ were found in an average of 8 seconds on a laptop computer.

Finally, at the discretion of the researcher is the option to develop instead an average coverage interval. While computationally intensive, the average coverage adjusted fiducial interval reduces interval lengths while still providing good coverage, and the computational cost of this adjusted interval would be much lower than the cost of adding subjects to clinical trials. The option of the exact or average adjusted intervals provide researchers great flexibility, depending on their prioritization of coverage and length, for small sample problems involving the linear combination of multinomial probabilities. 



\appendix

\section*{Appendix 1}
\subsection*{Demonstration of distribution function's adherence to fiducial requirements}

An example is presented to demonstrate why the distribution functions for $Y=\widehat{L}$, defined in Equations \ref{CDF2} and \ref{CDF1}, depend entirely and only on $L$, a requirement for the application of the fiducial approach. 

Consider the simple Scenario C from our simulation, where $\textbf{w} = (1,0,-1,0).$
Let
\begin{equation}\nonumber
MN_{1} \sim MN(\mathbf{p_{1}},n_{1}), \mathbf{p_{1}}=(p_{11},1-p_{11}) 
\end{equation}
\begin{equation}\nonumber
MN_{2} \sim MN(\mathbf{p_{2}},n_{2}), \mathbf{p_{2}}=(p_{21},1-p_{21}) 
\end{equation}
Then
\begin{equation}\nonumber
L=1(p_{11})+0(1-p_{11})-1(p_{12})+0(1-p_{12})=p_{11}-p_{12}
\end{equation}
Now consider our definition of the CDF for the lower bound given in Equation \ref{CDF2}.
\begin{equation}\label{CDF2b}
\begin{split}
 P(Y \ge y \mid L) & =  1 - \inf_{\mathbf{p}: \mathbf{p}%
^{\prime }\mathbf{w} \le L}\left\{F_{Y}(y^{*}\mid \mathbf{p})\right\} \\
& = 1 - \inf_{\mathbf{p}: \mathbf{p}%
^{\prime }\mathbf{w} \le L}\left\{P(Y\le y^{*} \mid \mathbf{p} ) \right\} \\
& = 1 - \inf \left\{P(Y\le y^{*} \mid \mathbf{p}: \mathbf{p}'\mathbf{w}\le L ) \right\} 
\end{split}
\end{equation}
The infimum of the set in Equation \ref{CDF2b} is clearly unique. Therefore, to show $P(Y \ge y \mid L)$ only depends on $L$ we need to show the values of the set $\left\{P(Y\le y^{*} \mid \mathbf{p}: \mathbf{p}'\mathbf{w}\le L ) \right\} $ only depend on $L$.

First, $y^{*}$ is fixed for observed $\widehat{y}$ and this sample space for $Y$ consists of all the possible values of $\left( \mathbf{x%
}\circ \mathbf{n}\right) ^{\prime }\mathbf{w}$ that result from $\mathbf{x}%
\in \mathcal{B}$ and is denoted $\mathcal{Y}=\{y=\left( \mathbf{x}\circ 
\mathbf{n}\right) ^{\prime }\mathbf{w}:\mathbf{x}\in \mathcal{B}\}$. Therefore, $y^{*}$ depends only on $\mathbf{w}$, $n_{1}$, and $n_{2}$.

Second, $\forall$ $L$,  $\exists$ $\mathbf{p}$ such that $\mathbf{p}'\mathbf{w}\le L$.Then let the set of all $\mathbf{p}$ such that $\mathbf{p}'\mathbf{w}\le L$, be denoted $\mathbf{p_{L}}$. All values of $\mathbf{p_{L}}$ depend on $L$ and $\mathbf{w}$ only, and result in a distinct value for $P(Y\le y^{*})$. 

Therefore, the set $\left\{P(Y\le y^{*} \mid \mathbf{p}: \mathbf{p}'\mathbf{w}\le L ) \right\} = \left\{P(Y\le y^{*} \mid \mathbf{p_{L}}) \right\} $ depends only on $\widehat{y}$, $n_{1}$, $n_{2}$, $\mathbf{w}$, and $L$, and $L$ is the only parameter. The same result is easily shown for for the CDF defined for the upper bound in Equation \ref{CDF1}.

\section*{Appendix 2}
\subsection*{Proof of adherence to exact coverage}

\begin{prop}
$y_{LB, \alpha/2}(L)$ defined in Equation \ref{yalphaLB} is non-decreasing in $L$.
\end{prop}
\begin{myproof}
Given the defintion of  $F_{Y, LB}(y \mid L)$ in Equation \ref{CDF2}, $L_{2} > L_{1} \implies F_{Y, LB}(y \mid L_{1}) \le F_{Y, LB}(y \mid L_{2})$.
Now, let $L_{2} > L_{1}$. Assume $y_{LB, \alpha}(L)$ is decreasing in $L$.  Then for any $\alpha \in (0,1)$,\\
$ y_{LB, \alpha}(L_{2} ) < y_{LB, \alpha}(L_{1})$ 
$\implies F_{Y, LB}(y_{LB, \alpha}(L_{2}) \mid L_{1}) > \alpha$ [Due to minimization in Equation \ref{yalphaLB} and $F_{Y, LB}(y \mid L)$ is decreasing in $y$ given summation in Equation \ref{pdf1}]. Also, \\
$\alpha \ge F_{Y, LB}(y_{LB, \alpha}(L_{2}) \mid L_{2}) \ge  F_{Y, LB}(y_{LB, \alpha}(L_{2}) \mid L_{1})$ [from Equation \ref{yalphaLB} and Preliminary].  \\
$\implies F_{Y, LB}(y_{LB, \alpha}(L_{2}) \mid L_{1}) \le \alpha$, $\Rightarrow\!\Leftarrow$,
$\therefore$ $y_{LB, \alpha/2}(L)$ is non-decreasing in $L$. 
\end{myproof}

 \begin{prop}$y_{UB, \alpha/2}(L)$ defined in Equation \ref{yalphaUB} is non-decreasing in $L$.
 \end{prop}
 \begin{myproof}
Given the defintion of  $F_{Y, UB}(y \mid L)$ in Equation \ref{CDF1}, $L_{2} > L_{1} \implies F_{Y, UB}(y \mid L_{1})\ge F_{Y, UB}(y \mid L_{2})$.\\
Now, let $L_{2} > L_{1}$. Assume $y_{UB, \alpha}(L)$ is decreasing in $L$.  Then for any $\alpha \in (0,1)$,
$ y_{UB, \alpha}(L_{2} ) < y_{UB, \alpha}(L_{1})$ 
$\implies F_{Y, UB}(y_{UB, \alpha}(L_{1}) \mid L_{2}) > \alpha$ [from maximization in Equation \ref{yalphaUB}  and $F_{Y, UB}(y \mid L)$ is increasing in $y$ given summation in Equation \ref{pdf1}]. Also, 
$\alpha \ge F_{Y, UB}(y_{UB, \alpha}(L_{1}) \mid L_{1}) \ge F_{Y, UB}(y_{UB, \alpha}(L_{1}) \mid L_{2})$ [from Equation \ref{yalphaUB} and Preliminary].  \\
$\implies F_{Y, UB}(y_{UB, \alpha}(L_{1}) \mid L_{2}) \le \alpha$ , $\Rightarrow\!\Leftarrow$,
$\therefore$ $y_{UB, \alpha/2}(L)$ is non-decreasing in $L$. 
\end{myproof}

\begin{prop} $L_{LB, \alpha}(y)$ and $L_{UB, \alpha}(y)$ give an interval with $1-\alpha$ or greater coverage on $L$. 
\end{prop}
\begin{myproof}
Following \citet{Pedersen1978}, for a \textcolor{black}{set of observations} $y_{1}, y_{2}, \dots$ on $Y$ corresponding to the parameters $L_{1}, L_{2}, \dots $ we have,
for the lower bound: \\
$P[y_{i} > y_{LB, \alpha / 2}(L_{i})] \le P[y_{i} \ge y_{LB, \alpha / 2}(L_{i})] = F_{Y, LB}(y_{LB, \alpha / 2}(L_{i}) \mid L_{i})  \le \alpha / 2$  from Equation \ref{yalphaLB}.
Then, by Proposition 1 and infimum and minimization in Equations \ref{LL} and \ref{yalphaLB}, respectively, $y_{i} > y_{LB, \alpha / 2}(L_{i}) \Leftarrow\!\Rightarrow L_{LB, \alpha / 2}(y_{i}) \ge L_{i}$ $\implies P[L_{i} < L_{LB, \alpha / 2}(y_{i})] = P[L_{i} \le L_{LB, \alpha / 2}(y_{i})]  \le P[y_{i} \ge y_{LB, \alpha / 2}(L_{i})]  \le \alpha / 2.$ \\
Additionally, 
 for the upper bound: \\
$P[y_{i} > y_{UB, \alpha / 2}(L_{i})] = 1 - P[y_{i} \le y_{UB, \alpha / 2}(L_{i})] = 1 - F_{Y,UB}(y_{UB, \alpha / 2}(L_{i}) \mid L_{i}) \ge 1 - \alpha / 2$ from Equation \ref{yalphaUB}.
Then, by Proposition 2 and the supremum and maximization in Equations \ref{LU} and \ref{yalphaUB}, respectively,
we have $y_{i} > y_{UB, \alpha / 2}(L_{i}) \Leftarrow\!\Rightarrow L_{UB, \alpha / 2}(y_{i}) \ge L_{i}$\\ 
$\implies P[L_{i} < L_{UB, \alpha / 2}(y_{i})] = P[L_{i} \le L_{UB, \alpha / 2}(y_{i})] = P[y_{i} > y_{UB, \alpha / 2}(L_{i})] \ge 1 - \alpha / 2.$ \\
Using the results for the upper and lower bound, \\
$P[L_{i} \in [L_{LB, \alpha / 2}(y_{i}), L_{UB, \alpha / 2}(y_{i})] \mid y_{i}] = 
1 - (P[L_{i} <  L_{LB, \alpha / 2}(y_{i})] + P[L_{i} > L_{\alpha / 2}(y_{i})])\ge  
1 - (\alpha / 2 + \alpha /2) = 
1 - \alpha .$
$\therefore L_{LB, \alpha / 2}(y)$ and $L_{UB, \alpha / 2}(y)$ give an interval on $L$ for an observed $y \in \mathcal{Y}$ with coverage $\ge 1 - \alpha$. 
\end{myproof}

\bibliographystyle{dcu}


\end{document}